# Elastic Pseudoturbulence in Polymer Solutions

Mithun Ravisankar[1] and Roberto Zenit[1]

[1]*School of Engineering, Brown University, 184 Hope St, Providence, RI 02912, USA*
(Dated: March 4, 2025)

We study the effects of polymer additives on pseudoturbulence induced by a swarm of bubbles rising in a quiescent fluid. We find that, beyond a critical polymer concentration, the energy spectra of velocity fluctuations in bubble-induced turbulence decay more steeply with respect to the wavenumber $k$. This new scaling is significantly steeper than the classical $k^{-3}$ scaling observed for bubbles in Newtonian fluids; it is independent of the gas volume fraction in the inertial limit and occurs within the length scales between the bubble wake length and the bubble diameter. Furthermore, we provide strong evidence that the presence of polymers enhances the coherence of the flow, highlighting the significant role of polymer additives in modifying the characteristics of pseudoturbulence.

Buoyancy-driven turbulent flows occur in many natural and industrial processes, such as in the upper ocean layer [1] and bubble column reactors [2] where bubbles play a pivotal role in inducing mixing. The addition of bubbles to turbulent flows has been studied for decades [3] owing to the significant drag reduction [4], heat transfer enhancement [5–7], among other effects [8, 9]. When a swarm of bubbles rise in an otherwise stagnant fluid, the disturbance caused by each bubble interacts with one another giving rise to velocity fluctuations in the continuous liquid phase. This fluctuating motion is referred to as bubble-induced turbulence, more commonly "pseudoturbulence" [10].

A paramount attribute of bubble-induced turbulence is its departure from the homogeneous isotropic Kolmogorov's turbulence, as summarized by Risso [11]. Lance and Bataille [3] first observed that the addition of bubbles to a developed Newtonian turbulence leads to an energy spectra of velocity fluctuations, $E(k) \sim k^{-3}$, where $k$ is the wavenumber, in the inertial subrange as opposed to the classical $k^{-5/3}$ decay observed in isotropic turbulence. This finding inspired numerous studies, both experimental [12–15] and numerical [16, 17], which supported the idea that the $k^{-3}$ scaling emerges in a homogeneous bubbly flow from both the spatial and temporal velocity disturbances. Specifically, the spatial contribution dominates at large scales, while the temporal contribution prevails at smaller scales [11].

It was previously argued that, in a statistically steady state, the $k^{-3}$ scaling arises from a spectral balance between viscous dissipation and energy production by bubbles [3]. However, coarse-grained simulations of pseudoturbulence [18], using a spectral decomposition of the energy budget, refuted this argument, showing that no scales exist where energy production and dissipation are in equilibrium. Building on this complexity, an intriguing characteristic of pseudoturbulence is the emergence of the flow agitations in the continuous liquid phase independent of whether the dispersed phase is composed of solid spheres [19] or bubbles [14] for a wide range of Reynolds number and Weber number [17, 20]. This characteristic $k^{-3}$ scaling arises from wake-wake interactions between rising bubbles, as confirmed by point particle simulations, which do not exhibit the $k^{-3}$ scaling due to the absence of wakes behind the bubbles [21]. Recent experimental investigations further confirmed that reducing the bubble Reynolds number (as the wake behind the bubble vanishes) replaces the $k^{-3}$ scaling observed in pseudoturbulence with a $k^{-5/3}$ scaling [22].

For certain applications like porous media flows [23], polymers are employed to achieve maximum drag reduction [24, 25] and heat transfer enhancement [26]. Because of the stretching property of the polymer molecules, efficient mixing can be achieved even in the absence of inertia (very small Reynolds number), referred to as "elastic turbulence" [27, 28]. When the Reynolds number is sufficiently high such that the inertial effects cannot be neglected, the modified turbulence due to the elasticity is referred to as elastoinertial turbulence [29]. A combination of experimental and numerical studies in the elastoinertial turbulence regime has revealed that a scaling exponent of $-3$, in contrast to the classical Kolmogorov's $-5/3$ scaling observed in Newtonian fluids, is a distinctive signature of polymer induced turbulence [30–33]. Importantly, the $-3$ scaling observed in Newtonian pseudoturbulence and that in polymer-induced turbulence are coincidental and arise from entirely different mechanisms.

While the modulation of turbulence spectra by polymers [28, 29, 33] and by the bubbles [3, 12, 14] has interested the scientific community for decades, an intriguing open question remains: how do polymer solutions modify the bubble-induced turbulence? In this Letter, we report evidence of a significant modulation in the turbulent energy spectra by a swarm of bubbles rising in quiescent aqueous polymer solutions. We argue that the eddies generated by the wake interactions of bubbles in polymer solutions, which differ significantly from those in the Newtonian fluids [34, 35], play a critical role on the modulation of turbulence spectra. Consequently, there is an enhancement of large-scale coherent structures consistent with the recent experimental [36] and simulation results [26]. This new found insight opens new avenues in the elastoinertial turbulence and has direct relevance for polymer-based drilling slurries used in oil and natural gas explorations [37].

In our experiments, mono-dispersed bubbly swarms were introduced into a quiescent aqueous polymer solution using a capillary bank [12]. We study the wake behind the bubble swarm using high-speed particle image velocimetry, by abruptly stopping the bubble formation with a solenoid valve, as proposed by [14]. The results for viscoelastic fluids, obtained by varying the mean gas volume fraction, $\alpha$, and the polymer concentration, are contrasted with those of Newtonian fluids, which serves as a benchmark. Details of the experimental setup and rheological characterizations are provided in the Supplemental Material [38]. We present our findings by considering the values of the relevant dimensionless groups: (i) Reynolds number, Re $= \rho U_b d/\eta$, where $U_b$ is the bubble velocity, $d$ is the bubble diameter, $\rho$ is the fluid density, and $\eta$ is the fluid viscosity, (ii) Weber number, We $= \rho U_b^2 d/\sigma$, where $\sigma$ is the surface tension, (iii) Weissenberg number, which measures the elasticity in the polymer solutions, is defined as, Wi $= \lambda_e U_b/d$, where $\lambda_e$ is the relaxation time, and (iv) Elasticity number, which compares the elastic stresses to the inertial stresses, is defined as, El = Wi/Re.

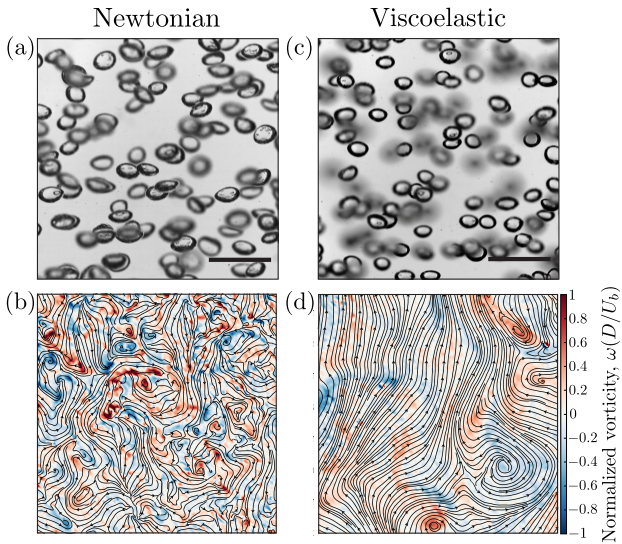

FIG. 1. Sample snapshots of bubbly flow in (a) Newtonian fluid (Re = 626, Wi = 0, El = 0) and (c) viscoelastic fluid (Re = 451, Wi = 3.8, El = 0.008) at a gas volume fraction of $\alpha \approx 0.025$. The scale bar represents 10 mm. Snapshots of vorticity, normalized by flow time $(d/U_b)$, from the PIV measurement behind the bubble swarm at the same gas volume fraction are shown for (b) Newtonian and (d) viscoelastic fluid. Solid lines indicate instantaneous streamlines in the wake.

Figure 1(a,c) show the snapshots of bubbly flow and Fig. 1(b,d) show the dimensionless vorticity fields, $\omega(d/U_b)$, obtained from the bubble swarm wake for the Newtonian fluid and viscoelastic fluid, respectively. We note that the differences in the flow field are not caused due to the difference in the Reynolds number considered. From the instantaneous streamlines overlaid on the normalized vorticity fields it is evident that the flow structures in the wake of the bubble swarm in the polymeric fluid is significantly different from that of the Newtonian fluid. We can quantify this observation considering the spatial correlation across a length scale $r$ of the horizontal velocity, $u_x$, and vorticity magnitude, $\omega$, defined as, $\overline{R}_{xx} = \langle u_x(\mathbf{x}) \cdot u_x(\mathbf{x}+\mathbf{r}) \rangle / \langle u_x^2 \rangle$ and $\overline{R}_\omega = \langle \omega(\mathbf{x}) \cdot \omega(\mathbf{x}+\mathbf{r}) \rangle / \langle \omega^2 \rangle$, respectively. Here, $r$ lies solely in the horizontal plane and the $\langle \rangle$ represent the average in space. Figures 2(a) and 2(b) show that the correlation of horizontal velocity and vorticity magnitude increases over longer distances as the polymer concentration increases, at a constant gas volume fraction of $\alpha \approx 0.025$, respectively. The average vortex size can be then estimated by the integral length scale of the vorticity, $L_\omega = \int_0^\infty \overline{R}_\omega \, dr$. Inset in the Fig. 2(b) shows that the average vortex size increases with the Weissenberg number. We note that the spatial correlations over the normalized distance does not change significantly with the gas volume fraction. Overall, this indicates that polymer stresses enhance the prominence of large scale flow structures, in agreement with previous experimental [39] and numerical simulations [26]. Here, the horizontal velocity is preferred over the velocity magnitude, because the probability density functions (PDFs) of the horizontal velocity fluctuations are symmetric, whereas the vertical velocity fluctuations are positively skewed (see Fig. S3 and S4 in the Supplemental Material [38]). In other words, there is an anisotropy in the vertical direction due to the mean motion of bubbles.

Now, we consider the second-order longitudinal velocity structure function defined as, $S_2(r) = \langle [u_x(\mathbf{x}+\mathbf{r}) - u_x(\mathbf{x})] \cdot (\mathbf{r}/r)^2 \rangle$ to explore on how the energy spectra will decay [40]. Here the $\langle \rangle$ denotes the spatial average. From the theory of homogeneous and isotropic turbulence for Newtonian incompressible fluids [41], we know that when the second-order velocity structure function scales $S_2(r) \sim r^\beta$, then the energy spectra should scale as $E(k) \sim k^{-(\beta+1)}$. Hence, for the classical Kolmogorov turbulence, the second-order velocity structure function behaves as $S_2(r) \sim r^{2/3}$ in the inertial range and $S_2(r) \sim r^2$ in the dissipative range. Figure 2(c) shows the second-order velocity structure function observed in the wake of the bubble swarm as a function of normalized length scale $r$ for a range of Weissenberg number at a constant gas volume fraction of $\alpha \approx 0.025$. In Newtonian pseudoturbulence, however, because of the bubble wake-wake interactions in the intermediate length scales, instead of the $S_2(r) \sim r^{2/3}$ scaling in the inertial range the second-order structure function scales as $S_2(r) \sim r^2$ [20, 42]. Note that the $S_2(r) \sim r^2$ scaling for the Newtonian pseudoturbulence in the intermediate range is identical to that of the scaling behaviour in the dissipative range for the classical Newtonian Kolmogorov turbulence. As the polymer concentration increases the second-order structure function $S_2(r)$ for pseudoturbulence shows a steepness close to $r^3$. However, this can be deceptive as any energy spectra $E(k) \sim k^{-\gamma}$ where $\gamma$ is greater than 3 will also show $r^2$ dependence in the second-order structure function, analogous to the two dimensional turbulence [43] and in some instances can only be identified by the higher order structure functions [44]. Thus for the viscoelastic pseudoturbulence we can conjecture that the energy spectra should decay as $E(k) \sim k^{-\gamma}$, where $\gamma$ is greater than 4.

Figure 3(a) and (b) show the energy spectra of horizontal and vertical velocity fluctuations normalized by their respective variance and bubble diameter for a range of

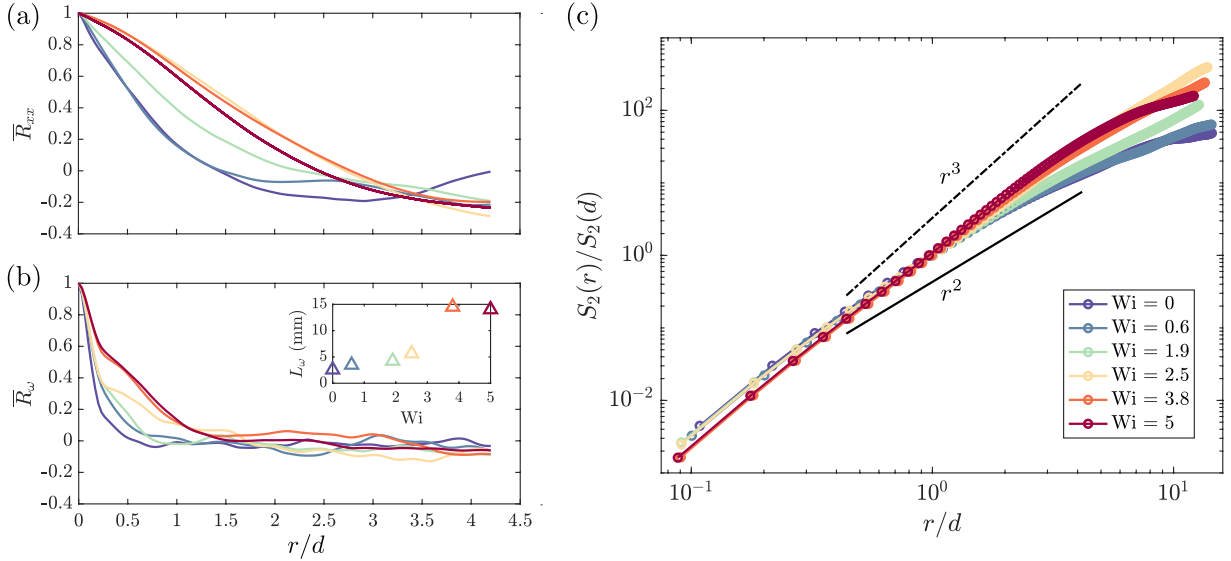

FIG. 2. Spatial correlation of (a) horizontal velocity and (b) vorticity magnitude in the bubble swarm wake for various Weissenberg numbers at a constant gas volume fraction ($\alpha \approx 0.025$). As the Weissenberg number increases, the flow gets correlation over longer distances. Inset: Average vortex size ($L_\omega$) vs. Weissenberg number. (c) Normalized second-order structure function, $S_2(r)$, vs. normalized length scale $r$ for different Weissenberg numbers. Solid and dot-dashed lines indicate $r^2$ and $r^3$ scaling, respectively.

Weissenberg numbers at a constant gas volume fraction of $\alpha \approx 0.025$. The details of the energy spectra calculations are provided in the Supplemental Material (see Fig. S5 and S6) [38]. Here the abscissa is normalised by the wavenumber corresponding to the bubble diameter, $k_d = 2\pi/d$. As is well established from decades of research [3, 11, 12, 14, 19, 45], for a bubbly flow in Newtonian fluid (Wi = 0), the energy spectra of the velocity fluctuations, $E(k)$, decays as $k^{-3}$, which is dubbed as the signature of pseudoturbulence for wavenumbers $k \leq k_d$. For different gas volume fractions, normalizing the energy spectra by the respective variance of the velocity fluctuations results in a single curve. This indicates that the velocity fluctuations increase with the gas volume fraction (See Fig. S7 in the Supplemental Material [38]).

A striking difference in the energy spectral decay is observed when the polymer concentration increases (Wi > 0). The $k^{-3}$ scaling of the energy spectra (for wavenumber $k \leq k_d$) is gradually replaced by a much steeper scaling of $E(k) \sim k^{-20/3}$ beyond a critical Weissenberg number, $\text{Wi}_{cr} \approx 2.5$. Such a steep exponent (approximately -20/3) has not been reported in the context of turbulence modulation by the elasticity. Similar to the Newtonian case, there is no change in the $E(k) \sim k^{-20/3}$ scaling in polymer solutions for a range of gas volume fractions (See Fig. S8 in the Supplemental Material [38]). This is in agreement with the previous works which showed that the energy spectra does not depend on the gas volume fraction or the bubble diameter [12, 14]. Note that, at scales smaller than the bubble diameter (for wavenumber $k \geq k_d$), the classical Kolmogorov's power-law decay is recovered for all cases, irrespective of the Weissenberg number. The compensated energy spectra shows this information in a slightly different manner (see Fig. S9 in the Supplemental Material) [38].

To understand the scaling observed in the pseudoturbulence of polymeric fluids, we consider the spectral decomposition of the energy balance,

$$\frac{d}{dt}E(k,t) = T(k,t) - T^{[p]}(k,t) - D(k,t) + P(k,t), \quad (1)$$

where $D$ represents the kinetic energy dissipation at a wavenumber $k$, $T$ is the kinetic energy transfer across wavenumber $k$ within the solvent, $T^{[p]}$ is the kinetic energy transfer across wavenumber $k$ between the solvent and polymer molecules, and $P$ is the energy production due to the rising bubbles, respectively. At statistically steady state, the left-hand side of the Eq. 1 becomes zero, thus,

$$T(k) - T^{[p]}(k) = D(k) - P(k). \quad (2)$$

We first focus on the case of Newtonian fluids (Wi = 0), meaning $T^{[p]}(k) = 0$. Using a scale-by-scale energy budget Zamansky *et al.* [18] showed that at large scales ($k/k_d < 0.4$), the rate of energy injected by the bubbles, $P(k)$, dominates over the viscous dissipation, $D(k) = 2\nu k^2 E(k)$, thus $P(k) \approx -T(k)$, leading to $E(k) \sim k^{-1}$. Whereas, at small scales ($k/k_d > 1$), viscous dissipation becomes dominant, thus $D(k) \approx T(k)$, resulting in the classical Kolmogorov scaling $E(k) \sim k^{-5/3}$. However, in the intermediate scales, contributions from all the terms are non-negligible, thus, $T(k) = D(k) - P(k)$. From numerous experiments and simulations [11, 13, 14, 16, 18, 19, 46], we know that the contribution from all these terms gives rise to $E(k) \sim k^{-3}$ in the intermediate scales, driven by the wake-wake interactions.

In the case of polymeric fluids (Wi > 0), the kinetic energy transfer across the wavenumber $k$ between the solvent and polymer molecules, $T^{[p]}(k) \neq 0$. From Fig.

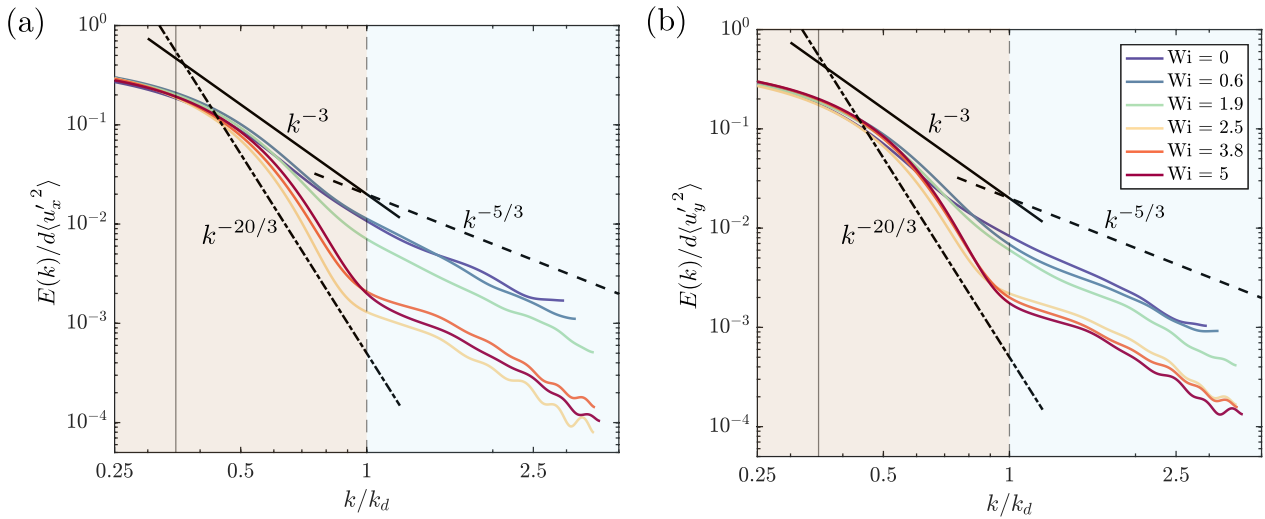

FIG. 3. (a) Horizontal and (b) vertical spectra of liquid velocity fluctuations, normalized by variance and bubble diameter, for various Weissenberg numbers at a constant gas volume fraction ($\alpha \approx 0.025$). The abscissa is normalized by the wavenumber corresponding to the bubble diameter ($k_d$). The solid and dashed gray lines indicate the upper and lower bounds of the $k^{-3}$ spectra in Newtonian fluids. The solid and dot-dashed black lines represent pseudoturbulence scaling for Newtonian ($k^{-3}$) and viscoelastic ($k^{-20/3}$) fluids at length scales relevant to the bubble wake and bubble diameter, respectively. The dashed black line denotes Kolmogorov scaling ($k^{-5/3}$) for both fluid types at scales smaller than the bubble diameter. Wake interaction and viscous dissipation regions ($k \leq k_d$ and $k \geq k_d$, respectively) are shaded light brown and light blue.

3, it is evident that even in the polymeric fluids, the energy spectra at large ($k/k_d \leq 0.4$) and at small scales ($k/k_d \geq 1$) behaves similar to that of the Newtonian case, leading to believe that the $T^{[p]}$ is important only in the intermediate scale. Note that the term $T^{[p]}$ is strictly positive in statistically steady state, as it is equal to the elastic energy dissipated by the polymers [47]. Because of this dissipative nature of the polymers (see Fig. S10 in the Supplemental Material [38]), the net effect of $T^{[p]}$ over all scales is to remove kinetic energy from the solvent and thus the negative sign in the Eq. 2 [48]. Thus, the contribution from the $T^{[p]}$ causes the energy spectra to decay more steeply than $k^{-3}$, leading to a scaling of $E(k) \sim k^{-\gamma}$, where $\gamma$ is greater than 4. This further confirms the scaling observed from the second-order structure function. From experiments we show that the $\gamma \approx 20/3$ for the elastic pseudoturbulence.

Motivated by recent work in relating the polymeric stress to the Lagrangian stretching fields [49], we calculate the Lagrangian coherent structures (LCS) [50, 51] to understand the effect of elasticity on the flow structures. In the present study, to further demonstrate the enhancement in the large scale turbulence over the small scale turbulence, the coherent structures are determined using finite time Lyapunov exponent (FTLE). In other words, when two states (or particles dispersed in a flow) separated by an infinitesimally small distance, $\delta_0$, the distance between these two states evolves as $\delta_0 e^{\lambda_L t}$, where $\lambda_L$ is the Lyapunov exponent. Thus, the FTLE field measures particle separation rate in the flow. This is helpful in understanding chaotic mixing of fluid elements over extended times [52]. Since FTLE is sufficient to locate the repelling and attracting LCS [53], the forward FTLE fields is determined using the TBarrier tool [54]. Calculating the FTLE of the flow fields in the bubble swarm wake provides insights into how a scalar field or Stokesian seeding particles would evolve [55]. Thus, with the velocity data known, the particle trajectories over time, $\mathbf{x}(t; t_0, \mathbf{x}_0)$, are determined from their initial position, $\mathbf{x}_0$ and initial time, $t_0$. The flow map, $\mathbf{F}_{t_0,t_1}(\mathbf{x}_0) = \mathbf{x}(t; t_0, \mathbf{x}_0)$ is defined from these particle trajectories. The FTLE field over the time interval $t_0$ to $t_1$ is defined as,

$$\text{FTLE}_{t_0,t_1}(\mathbf{x}_0) = \frac{1}{2|t_1 - t_0|} \log \lambda_{\max}(\mathbf{C}_{t_0,t_1}(\mathbf{x}_0)) \quad (3)$$

where $\lambda_{\max}$ is the largest eigenvalue of the right Cauchy-Green strain tensor, $\mathbf{C}_{t_0,t_1} = [\nabla \mathbf{F}_{t_0,t_1}]^{\text{T}} \mathbf{F}_{t_0,t_1}$. The time interval to determine the FTLE field is chosen to be the same dimensionless flow time $t^* = tU_b/d$ for all the fluids considered in the experiments. Figure 4(a) and (b) shows the normalized FTLE fields in the wake region behind the swarm of bubbles in Newtonian and viscoelastic fluid, respectively. In the contours of the normalized FTLE fields, a value of 0 corresponds to trenches while 1 corresponds to ridges. Ridges in the forward FTLE are associated with the regions of high shear i.e., the regions of intense mixing and stretching in the flow [52]. Compared to that of the Newtonian FTLE field, it is immediately clear that the presence of polymers significantly changes the regions of high shear. In other words, the small scale fine structures are smoothened in the presence of polymers. This confirms that the spatio-temporal evolution of the material regions in the viscoelastic fluid becomes more coherent at large scales (see Fig. S11 in the Supplemental Material) [56].

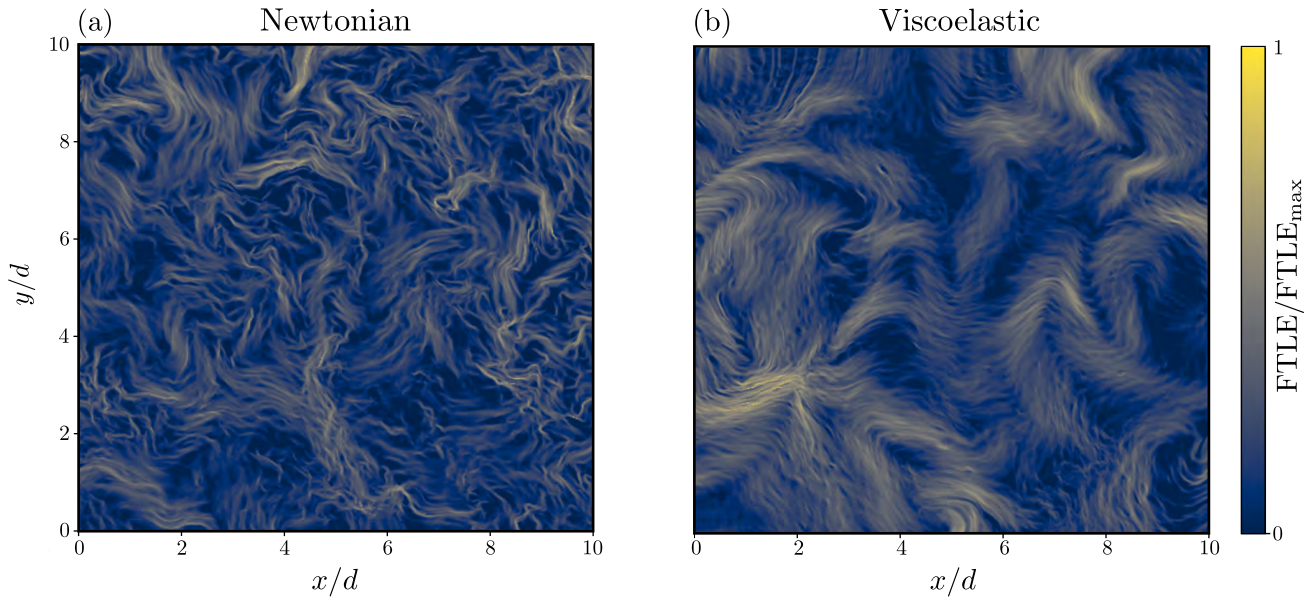

FIG. 4. Normalized FTLE field at the same dimensionless flow time interval, $t^* = 10$ right after the bubbles have left the measurement section in (a) Newtonian (Wi = 0) and (b) viscoelastic fluid (Wi = 3.8). In the contour, 0 corresponds to the trenches and 1 corresponds to the ridges. Notice the suppression of small scale structures and enhancement of large scale structures in the presence of polymer.

In this Letter, we have demonstrated for the first time that the addition of polymer additives significantly alters the bubble-induced turbulence, opening new avenues for investigating elasto-inertial turbulence [29]. Our findings suggest that the dissipative nature of the polymers results in a more rapid decay of the energy spectrum, exceeding the typical $k^{-3}$ scaling expected for the pseudoturbulence. As the polymer concentration is increased, the flow structure is correlated over longer distances and the energy spectra steepens from the classical $E(k) \sim k^{-3}$ and saturates at $E(k) \sim k^{-20/3}$ indicating an enhancement of energy transfer at intermediate length scales. Notably, the -20/3 exponent has not been observed for the turbulence of viscoelastic fluids to date; however, such a large energy decay for elastic pseudoturbulence lacks theoretical insights. Evidence from our experiments suggests that the -20/3 scaling exponent results from the synergistic contribution due to the presence of both bubbles and polymers. Given that bubbles in Newtonian fluids enhance heat transfer by nearly 20 times [6] and polymers alone increase heat transfer by 50% [26], the evidence presented here suggests that combining bubbles and polymer solutions could result in a dramatic enhancement of heat transfer efficiency.

*Acknowledgements* - The author M.R. thanks Prof. Martin Maxey, Prof. Frédéric Risso, Prof. Dominique Legendre and Prof. Varghese Mathai for crucial insights and stimulating discussions. M.R. thanks Sami Yamani Douzi Sorkhabi for helpful suggestions with the rheological characterisation. M.R. thanks Alex Pablo Encinas Bartos for helpful suggestions with the FTLE analysis.

[1] S. A. Thorpe and A. J. Hall, Nature **328**, 48 (1987).
[2] N. Kantarci, F. Borak, and K. O. Ulgen, Process Biochem. **40**, 2263 (2005).
[3] M. Lance and J. Bataille, J. Fluid Mech. **222**, 95 (1991).
[4] R. A. Verschoof, R. C. A. Van Der Veen, C. Sun, and D. Lohse, Phys. Rev. Lett. **117**, 104502 (2016).
[5] S. Dabiri and G. Tryggvason, Chem. Eng. Sci. **122**, 106 (2015).
[6] B. Gvozdić, E. Alméras, V. Mathai, X. Zhu, D. P. M. van Gils, R. Verzicco, S. G. Huisman, C. Sun, and D. Lohse, J. Fluid Mech. **845**, 226 (2018).
[7] O. Dung, P. Waasdorp, C. Sun, D. Lohse, and S. G. Huisman, J. Fluid Mech. **958**, A5 (2023).
[8] S. Balachandar and J. K. Eaton, Annu. Rev. Fluid Mech. **42**, 111 (2010).
[9] V. Mathai, D. Lohse, and C. Sun, Annu. Rev. Condens. Matter Phys. **11**, 529 (2020).
[10] L. Van Wijngaarden, Theor. Comput. Fluid Dyn. **10**, 449 (1998).
[11] F. Risso, Annu. Rev. Fluid Mech. **50**, 25 (2018).
[12] J. Martínez-Mercado, C. A. Palacios-Morales, and R. Zenit, Phys. Fluids **19**, 103302 (2007).
[13] E. Alméras, V. Mathai, D. Lohse, and C. Sun, J. Fluid Mech. **825**, 1091 (2017).
[14] G. Riboux, F. Risso, and D. Legendre, J. Fluid Mech. **643**, 509 (2010).
[15] V. N. Prakash, J. M. Mercado, L. van Wijngaarden, E. Mancilla, Y. Tagawa, D. Lohse, and C. Sun, J. Fluid Mech. **791**, 174 (2016).
[16] G. Riboux, D. Legendre, and F. Risso, J. Fluid Mech. **719**, 362 (2013).
[17] V. Pandey, R. Ramadugu, and P. Perlekar, J. Fluid Mech. **884**, R6 (2020).
[18] R. Zamansky, F. L. R. De Bonneville, and F. Risso, J. Fluid Mech. **984**, A68 (2024).

[19] Z. Amoura, C. Besnaci, F. Risso, and V. Roig, J. Fluid Mech. **823**, 592 (2017).
[20] T. Ma, H. Hessenkemper, D. Lucas, and A. D. Bragg, J. Fluid Mech. **936**, A42 (2022).
[21] I. M. Mazzitelli and D. Lohse, Phys. Rev. E. **79**, 066317 (2009).
[22] M. Ravisankar and R. Zenit, J. Fluid Mech. **1001**, A34 (2024).
[23] C. A. Browne and S. S. Datta, Sci. Adv. **7**, eabj2619 (2021).
[24] D. Samanta, Y. Dubief, M. Holzner, C. Schäfer, A. N. Morozov, C. Wagner, and B. Hof, Proc. Natl. Acad. Sci. **110**, 10557 (2013).
[25] G. H. Choueiri, J. M. Lopez, and B. Hof, Phys. Rev. Lett. **120**, 124501 (2018).
[26] G. Boffetta, A. Mazzino, S. Musacchio, and L. Vozella, Phys. Rev. Lett. **104**, 184501 (2010).
[27] A. Groisman and V. Steinberg, Nature **410**, 905 (2001).
[28] V. Steinberg, Annu. Rev. Fluid Mech. **53**, 27 (2021).
[29] Y. Dubief, V. E. Terrapon, and B. Hof, Annu. Rev. Fluid Mech. **55** (2023).
[30] W. Zhang, H. Zhang, Y. Li, B. Yu, and F. Li, Phys. Fluids **33** (2021).
[31] Y. Dubief, V. E. Terrapon, and J. Soria, Phys. Fluids **25** (2013).
[32] J. Song, F. Lin, N. Liu, X. Lu, and B. Khomami, J. Fluid Mech. **926**, A37 (2021).
[33] S. Yamani, B. Keshavarz, Y. Raj, T. A. Zaki, G. H. McKinley, and I. Bischofberger, Phys. Rev. Lett. **127**, 074501 (2021).
[34] R. Zenit and J. J. Feng, Annu. Rev. Fluid Mech. **50**, 505 (2018).
[35] M. Ravisankar, A. G. Correa, Y. Su, and R. Zenit, J. Non-Newtonian Fluid Mech. **309**, 104912 (2022).
[36] R. Vonlanthen and P. A. Monkewitz, J. Fluid Mech. **730**, 76 (2013).
[37] P. Beiersdorfer, D. Layne, E. W. Magee, and J. I. Katz, Phys. Rev. Lett. **106**, 058301 (2011).
[38] See the Supplemental Material [online](online) for details on experimental setup, rheological characterization, and supplemental figures.
[39] R. Ran, D. A. Gagnon, A. Morozov, and P. E. Arratia, arXiv preprint arXiv:2111.00068 (2021).
[40] Y. Zhang, E. Bodenschatz, H. Xu, and H. Xi, Sci. Adv. **7**, eabd3525 (2021).
[41] U. Frisch, *Turbulence: The Legacy of A.N. Kolmogorov* (Cambridge University Press, 1995).
[42] E. Trautner, M. Klein, F. Bräuer, and J. Hasslberger, Phys. Fluids **33** (2021).
[43] G. Boffetta and R. E. Ecke, Annu. Rev. Fluid Mech. **44**, 427 (2012).
[44] R. K. Singh, P. Perlekar, D. Mitra, and M. E. Rosti, Nat. Commun. **15**, 4070 (2024).
[45] F. Risso and K. Ellingsen, J. Fluid Mech. **453**, 395 (2002).
[46] I. Roghair, J. M. Mercado, M. V. S. Annaland, H. Kuipers, C. Sun, and D. Lohse, Int. J. Multiph. Flow. **37**, 1093 (2011).
[47] P. C. Valente, C. B. Da Silva, and F. T. Pinho, J. Fluid Mech. **760**, 39 (2014).
[48] P. C. Valente, C. B. da Silva, and F. T. Pinho, Phys. Fluids **28** (2016).
[49] M. Kumar, J. S. Guasto, and A. M. Ardekani, Proc. Natl. Acad. Sci. **120**, e2211347120 (2023).
[50] G. Haller, Annu. Rev. Fluid Mech. **47**, 137 (2015).
[51] G. Haller, *Transport Barriers and Coherent Structures in Flow Data* (Cambridge University Press, 2023).
[52] G. A. Voth, G. Haller, and J. P. Gollub, Phys. Rev. Lett. **88**, 254501 (2002).
[53] G. Haller and T. Sapsis, Chaos **21** (2011).
[54] A. P. Encinas Bartos, B. Kaszás, and G. Haller, [Zenodo](Zenodo) (2023).
[55] J. Peng and J. O. Dabiri, J. Fluid Mech. **623**, 75 (2009).
[56] Y. Dubief, J. Page, R. R. Kerswell, V. E. Terrapon, and V. Steinberg, Phys. Rev. Fluids **7**, 073301 (2022).

# Supplemental Material: Elastic Pseudoturbulence in Polymer Solutions

Mithun Ravisankar and Roberto Zenit
*School of Engineering, Brown University, 184 Hope St, Providence, RI 02912, USA*
(Dated: March 4, 2025)

**EXPERIMENTAL SETUP**

In our experiments, mono-dispersed air bubbles are injected into a stagnant fluid at the bottom of the tank through a bank of identical capillary needles of inner diameter 0.6 mm arranged in a hexagonal array as shown in Fig. S1. The mean gas volume fraction, $\alpha$ is controlled by adjusting the gas flow rate using a flowmeter. Here, the mean gas volume fraction is measured from the increase in the liquid level after the injection of bubbles, $\alpha = (H_0/\Delta H + 1)^{-1}$, where $H_0$ is the initial liquid level and $\Delta H$ is the liquid level increase. Measuring liquid velocity fluctuations in two-phase gas-liquid flows is particularly challenging due to the dispersed nature of bubbly flows. Conventional hot-wire-based techniques are often used [1–3], but they are imperfect, as they require filtering out portions of the signal where bubbles interact with the sensor. To overcome these difficulties, the liquid velocity fluctuations was measured by abruptly stopping the bubble formation using a solenoid valve. Once the airflow ceased, the bubbly flow was sharply cut off, allowing the last bubbles to cross the measurement section and leaving the wake region free of bubbles. Then, the wake behind the bubble swarm was studied using high-speed particle image velocimetry (Photron FASTCAM SA5 at 500 frames per second). Since the measurement is carried out in the wake region (after the passage of the bubbles), there was no need to use the fluorescent particles. A green laser beam (532 nm) from an Nd:YAG laser system was used for illuminating the 55 $\mu$m diameter tracer particles in the bubble column. The recorded images were then analyzed using PIVLab in MATLAB. The field of view is 38.5 mm × 38.5 mm (1024 × 1024 pixels). For the PIV analysis, 32 × 32 pixels interrogation regions and 50% overlap is used in the first pass and 16 × 16 pixels interrogation regions with 50% overlap on the subsequent pass is used. Each snapshot is composed of 127 × 127 vectors. The spatial resolution (physical distance between two neighbouring vectors) is 0.3 mm. Spurious vectors are detected by median test and replaced by interpolating neighbor vectors.

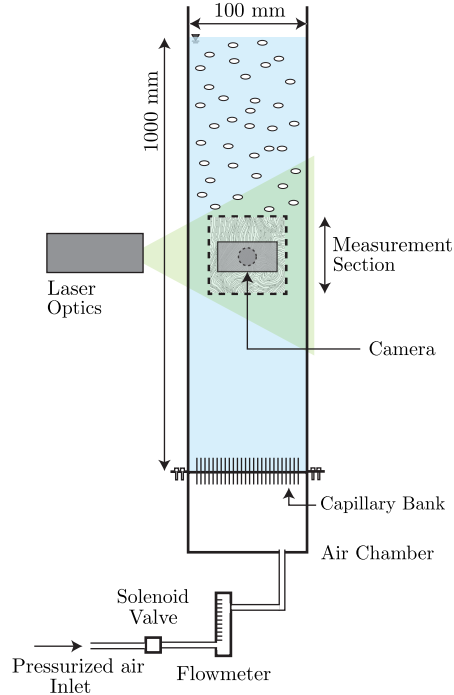

FIG. S1. Schematic of the experimental setup used in the current study showing the abruptly stopped injection of air bubbles with a solenoid valve to record the flow fields behind the bubble swarm using PIV. The size of the measurement section is 38.5 mm × 38.5 mm, depicted by a dashed box.

## RHEOLOGY OF FLUIDS

Aqueous solutions of non-ionic polyacrylamide polymer (PAAm) (Sigma Aldrich) with a molecular weight, $M_w$ of $5 \times 10^6$ g/mol were prepared by diluting a stock solution, which was prepared by dissolving polymer in deionized water. The solutions were gently mixed using an overhead mixer for at least 24 hrs to achieve homogeneity. To minimize bubble-bubble coalescence, a small amount of magnesium sulfate salt (0.025 mol/l) was added to all the solutions. The shear viscosity, density and surface tension of the fluids were measured using ARES-G2 Rheometer (TA Instruments), Density meter (Anton Paar) and Bubble pressure tensiometer (KRUSS Scientific Instruments), respectively. The properties of all the fluids used in the experiments are listed in Table S1. The intrinsic viscosity, $[\eta]$, is obtained using the Mark-Houwink equation, $[\eta] = \kappa M_w^a$. For the aqueous polyacrylamide solutions used in the experiments, the intrinsic viscosity is determined to be $[\eta] = 10.57$ dl/g, here the constants $\kappa = 10^{-4}$ dl/g and $a = 0.75$ [4]. The solvent to solution viscosity ratio is given by $\eta_{sp} = \eta_s/\eta$, where $\eta_s = 1.1$ mPa.s is the solvent viscosity. The coil-overlap concentration ($c^* = 1000$ ppm), was inferred from log-log plot of the zero-shear viscosity, $\eta_0$, against a wide range of polymer concentration, $c$ as shown in the Fig. S2(a). As the polymer concentration increases, the zero-shear viscosity changes the functional dependence from $\eta_0 \propto c^{0.6}$ to $\eta_0 \propto c^{3.3}$ clearly showing the non-linear polymer-polymer chain interactions [5]. To minimize the polymer-polymer chain interactions, we only consider fluids for which the polymer concentration, $c$, is well below the coil-overlap concentration, $c^*$. The shear-thinning observed in the dilute regime has a power index of $n \approx 0.95$ and hence ignored, following previous experimental studies [6, 7].

| Fluids (in water) | $c/c^*$ | $\rho$ (kg/m$^3$) | $\sigma$ (mN/m) | $\eta$ (mPa.s) | $\eta_{sp}$ | $\lambda_e$ (ms) | $d$ (mm) | $U_b$ (mm/s) | Re | We | Wi | El |
|---|---|---|---|---|---|---|---|---|---|---|---|---|
| 10% Glycerin | - | 1003.0 | 73.23 | 1.30 | - | $\approx 0$ | $2.8 \pm 0.2$ | 290 | 626 | 3.2 | 0 | 0 |
| 50 ppm PAAm | 0.05 | 1000.7 | 74.52 | 1.13 | 0.97 | 8.15 | $3.1 \pm 0.2$ | 224 | 614 | 2.1 | 0.6 | 0.001 |
| 100 ppm PAAm | 0.1 | 1001.4 | 74.35 | 1.27 | 0.86 | 29.1 | $3.4 \pm 0.1$ | 222 | 595 | 2.3 | 1.9 | 0.003 |
| 150 ppm PAAm | 0.15 | 1001.5 | 74.44 | 1.48 | 0.74 | 40.6 | $3.5 \pm 0.1$ | 218 | 502 | 2.2 | 2.5 | 0.005 |
| 200 ppm PAAm | 0.2 | 1001.8 | 74.50 | 1.65 | 0.66 | 62.7 | $3.5 \pm 0.2$ | 212 | 451 | 2.2 | 3.8 | 0.008 |
| 300 ppm PAAm | 0.3 | 1002.6 | 74.41 | 2.12 | 0.51 | 86.8 | $3.5 \pm 0.2$ | 206 | 340 | 2.0 | 5.0 | 0.015 |

TABLE S1. Physical properties of the fluids: $c$ - polymer concentration, $c^*$ - coil-overlap concentration; $\rho$ - density; $\sigma$ - surface tension; $\eta$ - viscosity; $\eta_{sp}$ - viscosity ratio; $\lambda_e$ - relaxation time; $d$ - bubble diameter; $U_b$ - bubble velocity; Re - Reynolds number; We - Weber number; Wi - Weissenberg number; El - Elasticity number.

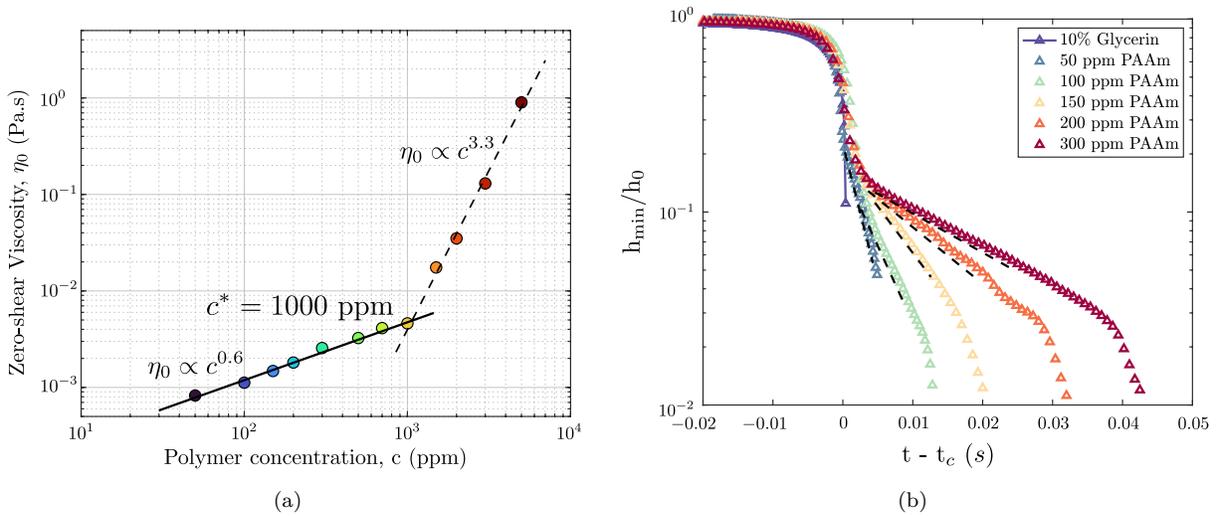

FIG. S2. (a) Zero-shear viscosity versus a wide range of polymer concentration for aqueous solutions of PAAm. The intersection of solid and dashed lines determine the coil-overlap concentration for PAAm solutions; (b) Evolution of minimum neck radius, $h_{\min}$, normalized by the nozzle radius, $h_0$, with the time axis shifted by the transition points, $t_c$, of different polymer concentration fluids. The relaxation time, $\lambda_e$, is determined from the slope of the dashed lines for each polymer concentration.

The relaxation time, $\lambda_e$, is determined using a dripping-onto-substrate method, following Dinic *et al.* [8]. For this, a pendant drop of fluid is brought into contact with a sessile drop on a glass substrate. This causes an unstable liquid bridge to form between the nozzle (of diameter $2h_0$) and the sessile drop, undergoing a capillary-driven thinning. By visualizing the collapse at 5000 frames per second, the liquid bridge undergoing inertio-capillary (observed before t - $t_c$ = 0) and elasto-capillary thinning dynamics (linear regime observed after t - $t_c$ = 0) can be determined from the subpixel image analysis using MATLAB as shown in the Fig. S2(b). Here, $t_c$ marks the transition from the inertio-capillary to elasto-capillary regime [8]. The relaxation time for varying polymer concentration is obtained from elastocapillary thinning dynamics described using

$$\frac{h_{\min}(t)}{h_0} \approx \left(\frac{Gh_0}{2\sigma}\right)^{1/3} \exp[-(t-t_c)/3\lambda_e], \tag{1}$$

where $G$ is the elastic modulus.

## PROBABILITY DENSITY FUNCTION OF THE VELOCITY FLUCTUATIONS

We now show the statistics of the velocity fluctuations and how they were altered in the presence of polymers. From previous studies [9–11], it has been shown that the PDFs of the horizontal velocity fluctuations are symmetric and non-Gaussian, whereas for the vertical component, the PDFs are strongly positively skewed (i.e. enhanced probability of upward fluctuations). This is because the positive skewness is the result of the velocity fluctuations in the vicinity of the bubble, which move vertically [9, 10]. Therefore, to capture these effects accurately, it is essential to analyze the velocity field within the bubble swarm rather than in the wake of the bubble swarm.

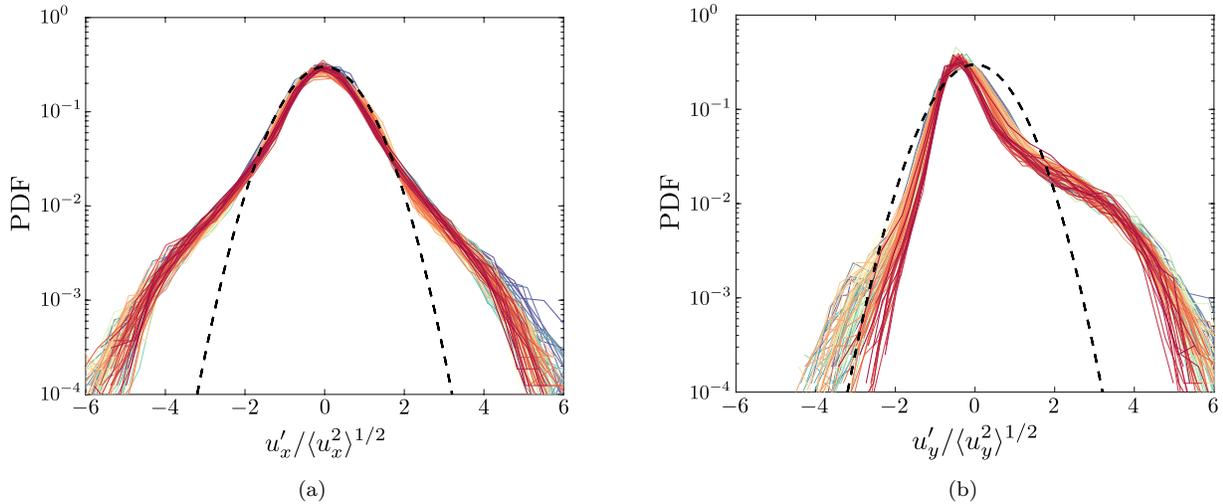

FIG. S3. PDF of the liquid velocity fluctuations (a) in the horizontal and (b) vertical direction within the bubbly flows in Newtonian fluid (Wi = 0) normalized by the standard deviation for $\alpha = 0.010$. The dashed line corresponds to the Gaussian profile. Here, the color gradients, ranging from blue to red, correspond to 2 ms of measurement within the bubble swarm

For that purpose, similar to [12, 13], fluorescent particles within the bubble swarm were added to capture the velocity fluctuations produced in the bubble vicinity. By nature, this particular visualization technique within the bubble swarm is restricted to low gas volume fractions. Figure S3 shows the PDF of the horizontal and vertical velocity fluctuations within the bubble swarm normalized by the standard deviation for gas volume fraction of $\alpha = 0.010$ for a Newtonian fluid. The multiple colors correspond to the consecutive instances (blue corresponds to $t = 0$ and red corresponds to $t = 2$ ms) in the decaying agitations left behind the bubble swarm. Both the horizontal and vertical velocity fluctuations PDFs are in good quantitative agreement with the results of Riboux *et al.* [10] who showed that PDFs corresponding to the flow in the vicinity of the bubbles was positively skewed whereas the PDFs corresponding to the wake of the bubble swarm was Gaussian. This was further corroborated by Alméras *et al.* [3] from the conditioned PDFs of a single bubble in a turbulent flow showing strong skewness in the primary and secondary wake whereas the conditioned PDFs in the far field are Gaussian (See Fig. 14(d) in [3]). Thus, the exponential tail is mainly due to the agitations by the wakes from the rising bubbles and their interactions, leading to a larger probability of upward fluctuations. For the case of viscoelastic fluids, as shown in the Fig. S4, the PDFs of the horizontal and vertical velocity fluctuations are similar in trend to that of the Newtonian fluids. The main difference is that there is a shrinkage in the exponential tail as observed in the case of Newtonian fluid. This can be interpreted as the polymer additives suppressing the exponential tails of velocity distributions by decreasing the occurance of large scale velocity fluctuations [14].

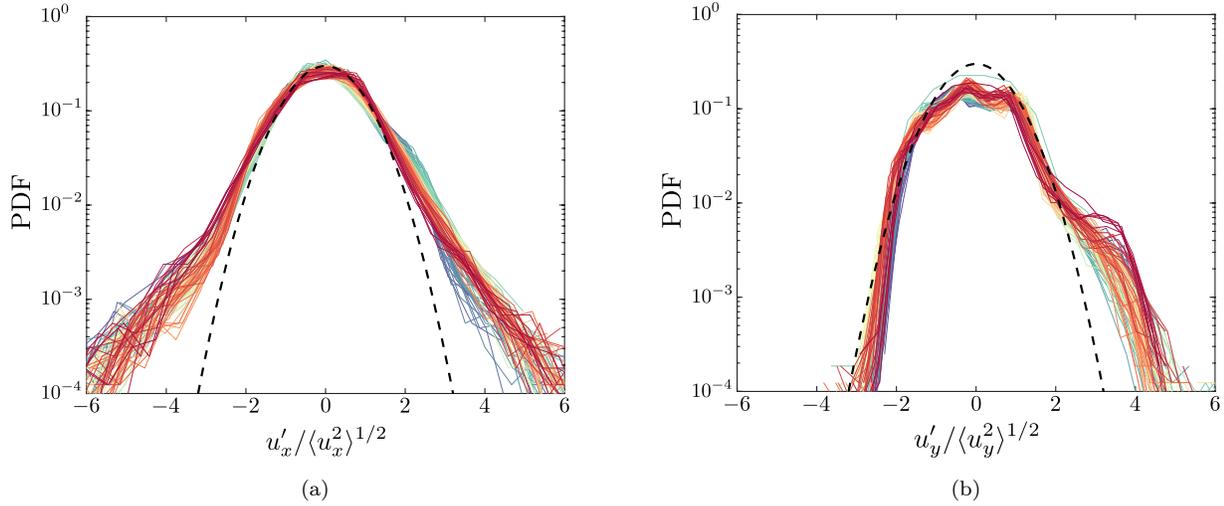

FIG. S4. PDF of the liquid velocity fluctuations (a) in the horizontal and (b) vertical direction within the bubbly flows in viscoelastic fluid (Wi = 3.8) normalized by the standard deviation for $\alpha = 0.010$. The dashed line corresponds to the Gaussian profile. Here, the color gradients, ranging from blue to red, correspond to 2 ms of measurement within the bubble swarm

## POWER SPECTRA OF VELOCITY FLUCTUATIONS

In the current experiments, the energy spectra of horizontal velocity fluctuations is calculated using the Welch method for each horizontal row of horizontal velocity fluctuations obtained from the particle image velocimetry (PIV) of the wake behind the bubble swarm, as detailed in [10]. The average energy spectra of horizontal velocity fluctuation is then obtained by taking the mean of energy spectra of all the rows, as shown in Fig. S5.

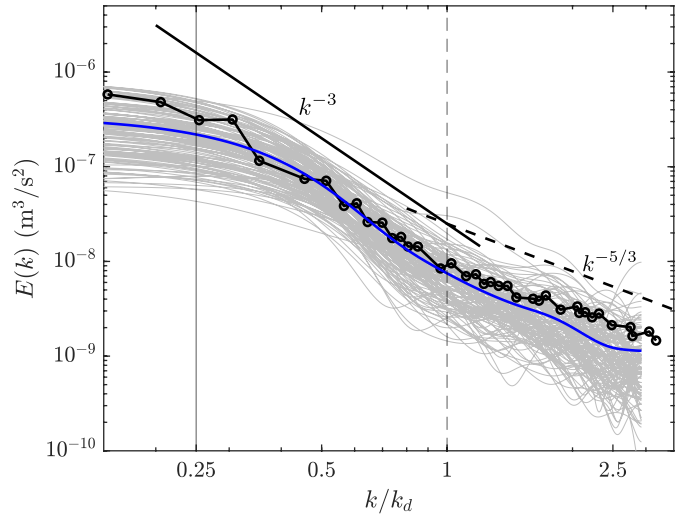

FIG. S5. Spatial spectrum of the horizontal velocity for each horizontal row of the PIV velocity field for the Newtonian fluid at Re = 626 and gas volume fraction, $\alpha \approx 0.0025$ is shown by the gray lines. The solid blue line corresponds to the average energy spectra of the horizontal velocity fluctuations. The abscissa is normalized by the wavenumber corresponding to the bubble diameter, $k_d$. We overlay the spectrum obtained from experiments by Riboux *et al.* [10] (solid black line with symbols) for Re = 670 and gas volume fraction, $\alpha \approx 0.0046$ and find it to be in agreement with our experiments. The solid and the dashed gray line in the abscissa denote the upper and lower boundaries of the $k^{-3}$ spectra observed by [10].

The average energy spectra of vertical velocity fluctuations in the vertical direction is obtained in the same manner as mentioned above. As established by Riboux *et al.* [10, 15], we have replicated the $k^{-3}$ scaling is observed between the Eulerian integral length scale ($\Lambda = d/C_d$) and the bubble diameter ($d$) in our benchmark Newtonian fluid bubbly flow experiments, where $C_d = \frac{16}{\text{Re}}\kappa(\text{Re})$ is the drag coefficient of a single rising bubble obtained from [16]. Here $\kappa(\text{Re}) = \frac{16+3.315\text{Re}^{1/2}+3\text{Re}}{16+3.315\text{Re}^{1/2}+\text{Re}}$. We note that the $k^{-3}$ scaling, observed less than a decade in the wavenumber space, is a defining feature of pseudoturbulence. As discussed in the main manuscript, pseudoturbulence originates from the wake interactions between rising bubbles. As a result, its characteristics are limited to the spatial range between the bubble wake length and the bubble diameter. In wavenumber space, $k^{-3}$ scaling is observed when $k \leq k_d$. Note that, at scales smaller than the bubble diameter (wavenumber $k > k_d$), the classical Kolmogorov's power-law decay, $E(k) \sim k^{-5/3}$ is recovered, again, in agreement with previous studies [10, 13, 17].

### TIME DEPENDENCE ON THE SPECTRA OF THE VELOCITY FLUCTUATIONS

Figure S6(a) shows the energy spectra of the decaying agitations measured after the bubbles have been switched off. For brevity only the horizontal energy spectra of the liquid velocity fluctutations are presented. It is evident that as the time increases (from blue to red), the agitation decays, hence the energy associated decreases; however, the shape of the curve remains unchanged. In other words, if the spectra is normalized by the corresponding variance, the spectra collapse into a single curve as shown in the Fig. S6(b). Therefore, to be consistent across the different experimental conditions, the flow is measured immediately after the bubble swarm has left the measurement section, where the energy associated is greater.

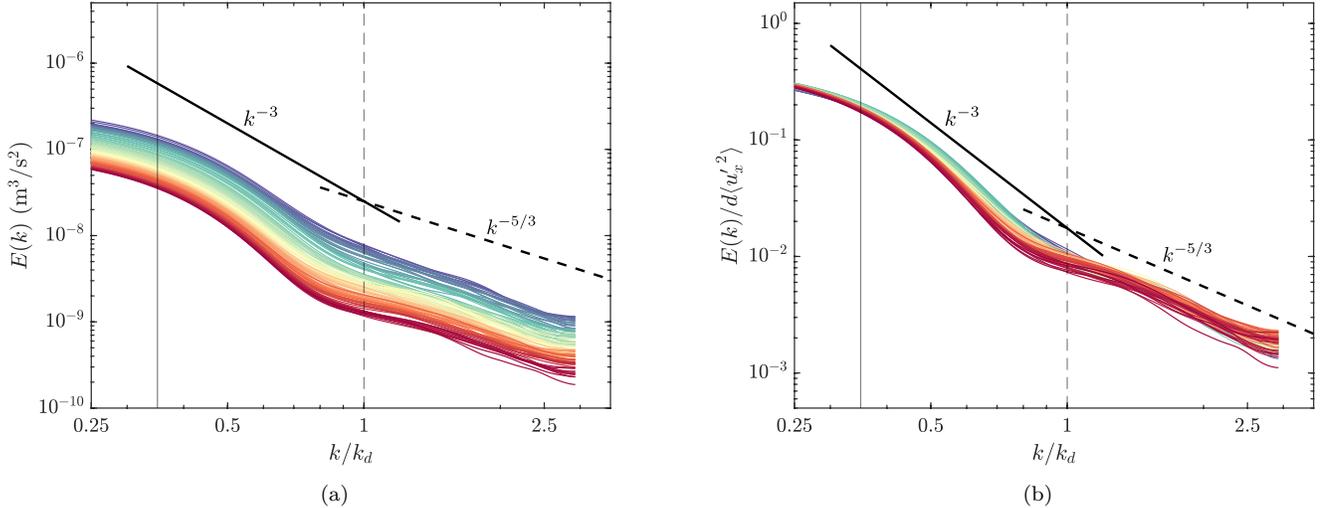

FIG. S6. (a) Energy spectra and (b) normalized energy spectra of the horizontal liquid velocity fluctuations with Re = 626 and gas volume fraction, $\alpha \approx 0.025$. The multiple colors correspond to the consecutive instances in the decaying agitations left behind the bubble swarm (blue corresponds to $t = 0$ and red corresponds to $t = 20$ ms). The $k^{-3}$ scaling in the intermediate scales (solid black line) and $k^{-5/3}$ scaling in the small scales (dashed black line) in accordance to [18].

### VARIANCE OF THE VELOCITY FLUCTUATIONS

Figure S7(a) shows the variance of the liquid velocity fluctuations as a function of gas volume fraction for a range of Weissenberg numbers. We first focus on the case of Newtonian fluids (Wi = 0). From their experiments, Alméras *et al.* [3] showed that the evolution of liquid agitation in bubbly turbulent flows depends on the so-called critical bubblance parameter, $b_c$. Here, the bubble parameter, which compares the energy of fluctuations produced by the bubble swarm to the energy of fluctuations produced by the incident turbulence, is given by,

$$b = \frac{V_r^2 \alpha}{{u_0'}^2}, \tag{2}$$

where, $u_0'$ is the incident turbulent fluctuations produced in the absence of bubbles (single-phase flows) and $V_r$ is the relative rising velocity of the bubble. They showed that regardless of the incident turbulent fluctuations ($u_0'$), the standard deviations of the liquid velocity fluctuations ($u_\mathrm{rms}$) showed a non-monotonic evolution with the bubblance parameter, $b$. The conclusions from their work is that the normalized velocity fluctuations, $u_\mathrm{rms}/u_0'$ evolves as $b^{0.4}$ when $b < b_c$, where $b_c \approx 0.7$ in their experiments. Whereas for $b > b_c$, the normalized velocity fluctuations increases much faster as $b^{1.3}$.

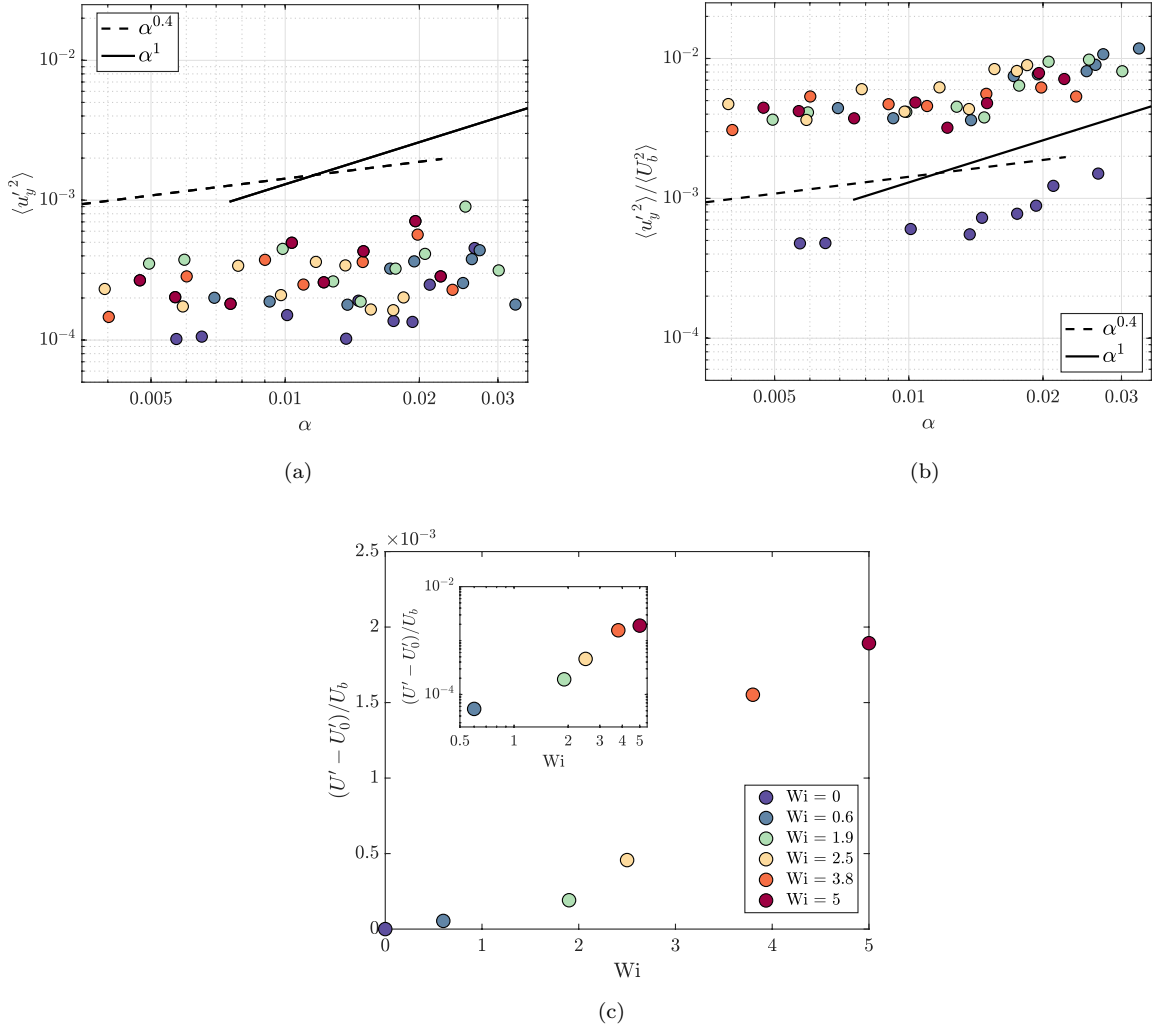

FIG. S7. (a) Variance of the velocity fluctuations, $\langle {u_y'}^2 \rangle$, in the vertical direction as a function of the gas volume fraction, $\alpha$, for a family of Weissenberg numbers; (b) Variance of the velocity fluctuations, $\langle {u_y'}^2 \rangle$, in the vertical direction normalized by the square of the mean bubble velocity, $\langle U_b^2 \rangle$, as a function of the gas volume fraction, $\alpha$, for a family of Weissenberg numbers. The dashed and solid lines show functional dependencies of $\langle {u_y'}^2 \rangle \propto \alpha^{0.4}$ and $\langle {u_y'}^2 \rangle \propto \alpha^1$, respectively; (c) Difference in the normalized RMS velocity fluctuations relative to the Newtonian case as a function of the Weissenberg number at a constant gas volume fraction of $\alpha \approx 0.02$; Inset shows the same plot in the log-log axis.

Note that $b = 0$ corresponds to single-phase flow, whereas $b = \infty$ corresponds to a bubble swarm rising in a quiescent liquid (pseudo-turbulence). In the current experiments, since the bubbles are rising in a quiescent liquid ($b = \infty$), we choose to compare the results using the gas volume fraction, $\alpha$. Figure S7(a) shows the variance of the velocity fluctuations as a function of gas volume fraction for a family of Weissenberg numbers. The velocity fluctuations are determined by $\mathbf{u}' = \mathbf{u} - \langle \mathbf{u} \rangle$. Here $\langle \rangle$ represent the average in space. For brevity, only the vertical velocity fluctuations are shown. It is immediately evident that the variance of the velocity fluctuations for a family of Weissenberg numbers are of the same order. Following the work of Cartellier and Rivière [19], the variance of the velocity fluctuations is normalized by the bubble velocity squared (see Fig. S7(b)). Similar to the previous works ([1, 3, 19]), for the Newtonian fluid (Wi = 0) two regimes can be observed, separated by a critical gas volume fraction of $\alpha_c \approx 0.015$. For $\alpha < \alpha_c$, the normalized velocity fluctuations increase monotonically as $\langle u'^2_y \rangle / \langle U_b^2 \rangle \propto \alpha^{0.4}$ (dashed line). Whereas for $\alpha > \alpha_c$, the normalized velocity fluctuations increase much faster as $\langle u'^2_y \rangle / \langle U_b^2 \rangle \propto \alpha^1$ (solid line). A similar trend is observed for the case of viscoelastic fluids on increasing the Weissenberg number. However, the normalized velocity fluctuations are an order of magnitude higher compared to that of the Newtonian case. This shows that the turbulence intensity is larger in the presence of elastic effects. This is further quantified using the root-mean-square (RMS) of the velocity fluctuations, given by $U' = \sqrt{(2\langle u_x^2 \rangle + \langle u_y^2 \rangle)/3}$. As shown in Fig. S7(c), the difference in the normalized RMS velocity fluctuations relative to the Newtonian case, $U'_0$, increases with the Weissenberg number at a constant gas volume fraction $\alpha \approx 0.02$, consistent with the previous studies [20, 21]. A clear transition from the Newtonian case occurs at a critical value of Weissenberg number, $\text{Wi}_{\text{cr}} \approx 2.5$.

**DEPENDENCE OF GAS VOLUME FRACTION ON THE SPECTRA OF VELOCITY FLUCTUATIONS**

For a Newtonian fluid, the average energy spectra, when normalized by the variance of the velocity fluctuations and bubble diameter, collapses into a single curve for different gas volume fractions as shown in the Fig. S8(a,b).

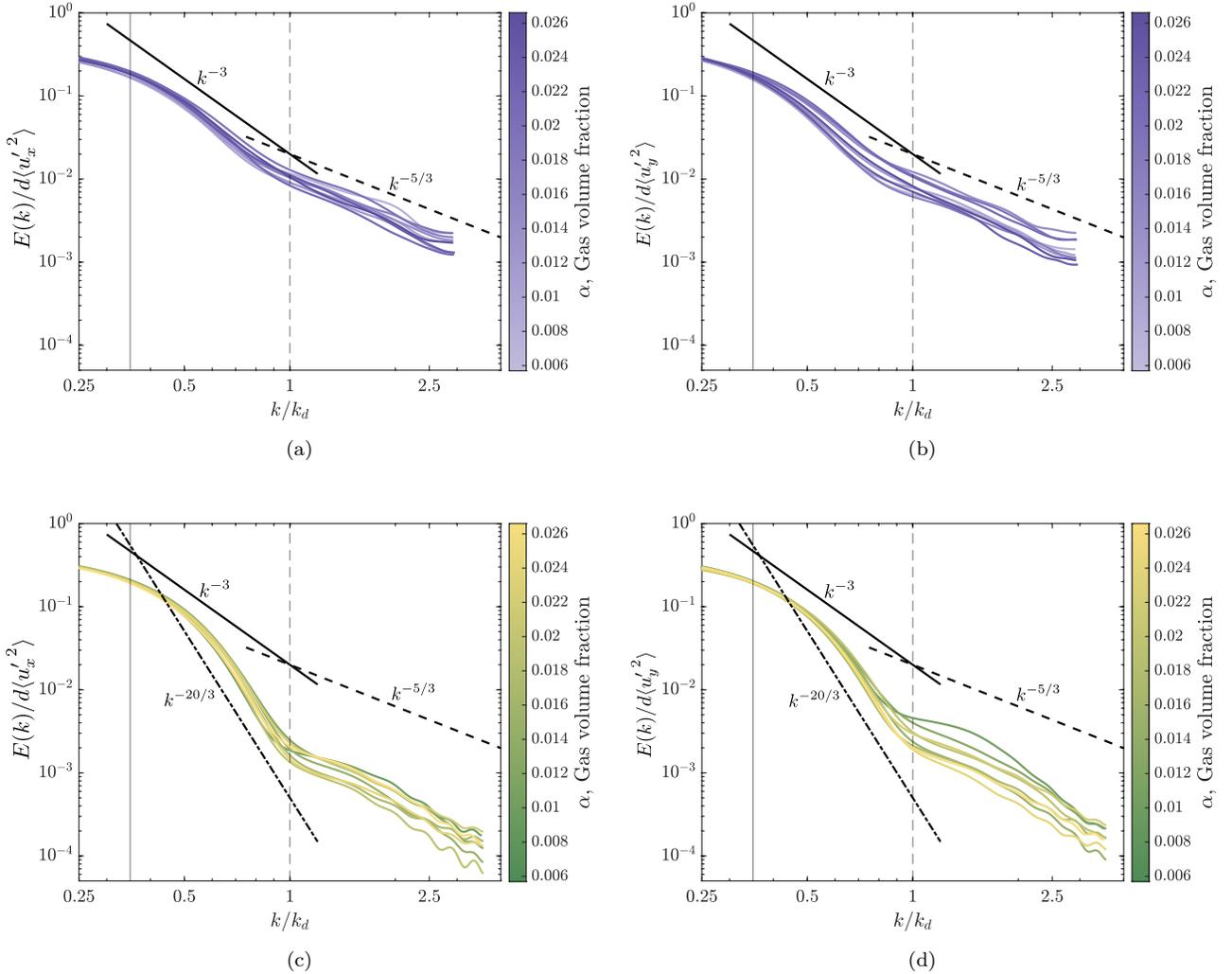

FIG. S8. Horizontal, and vertical spectra of the liquid velocity fluctuations in (a,b) Newtonian fluid and in (c,d) viscoelastic fluid (Wi = 3.8) normalized by the variance and bubble diameter for a family of gas volume fraction. The abscissa is normalized by the wavenumber corresponding to the bubble diameter, $k_d$. The solid and the dashed grey line in the abscissa denote the upper and lower boundaries of the $k^{-3}$ power spectral decay.

It is to be noted that, unlike the reported temporal power spectral density (refer to [3, 11]), the energy spectra obtained from the PIV images show a $k^{-3}$ scaling at large scales for less than a decade in the horizontal axis. Since the pseudoturbulence $k^{-3}$ scaling is a result of wake-wake interaction, the spatial energy spectra is expected to be observed between the bubble wake length and the bubble diameter ($k \leq k_d$). We report that in the viscoelastic fluids, similar to the Newtonian counter part, the average energy spectra when normalized by the variance of the velocity fluctuations and bubble diameter collapses into a single curve for a family of gas volume fractions with a steeper $k^{-20/3}$ scaling instead of the $k^{-3}$ scaling as shown in the Fig. S8(c,d).

## COMPENSATED ENERGY SPECTRA OF VELOCITY FLUCTUATIONS

Figure S9 shows the compensated energy spectra of the (a) horizontal and (b) vertical liquid velocity fluctuations showing the emergence of $k^{-3}$ scaling for a range of Weissenberg numbers at a constant gas volume fraction $\alpha \approx 0.025$, respectively. The abscissa is normalized by the wavenumber corresponding to the bubble diameter, $k_d$. We note that for Wi < Wi$_{cr}$, the $k^{-3}$ scaling is observed for wavenumbers, $k < k_d$. Furthermore, it is evident from the Fig. S9(c) and (d) that for Wi $\geq$ Wi$_{cr}$, the $k^{-20/3}$ scaling is observed for wavenumbers, $k < k_d$.

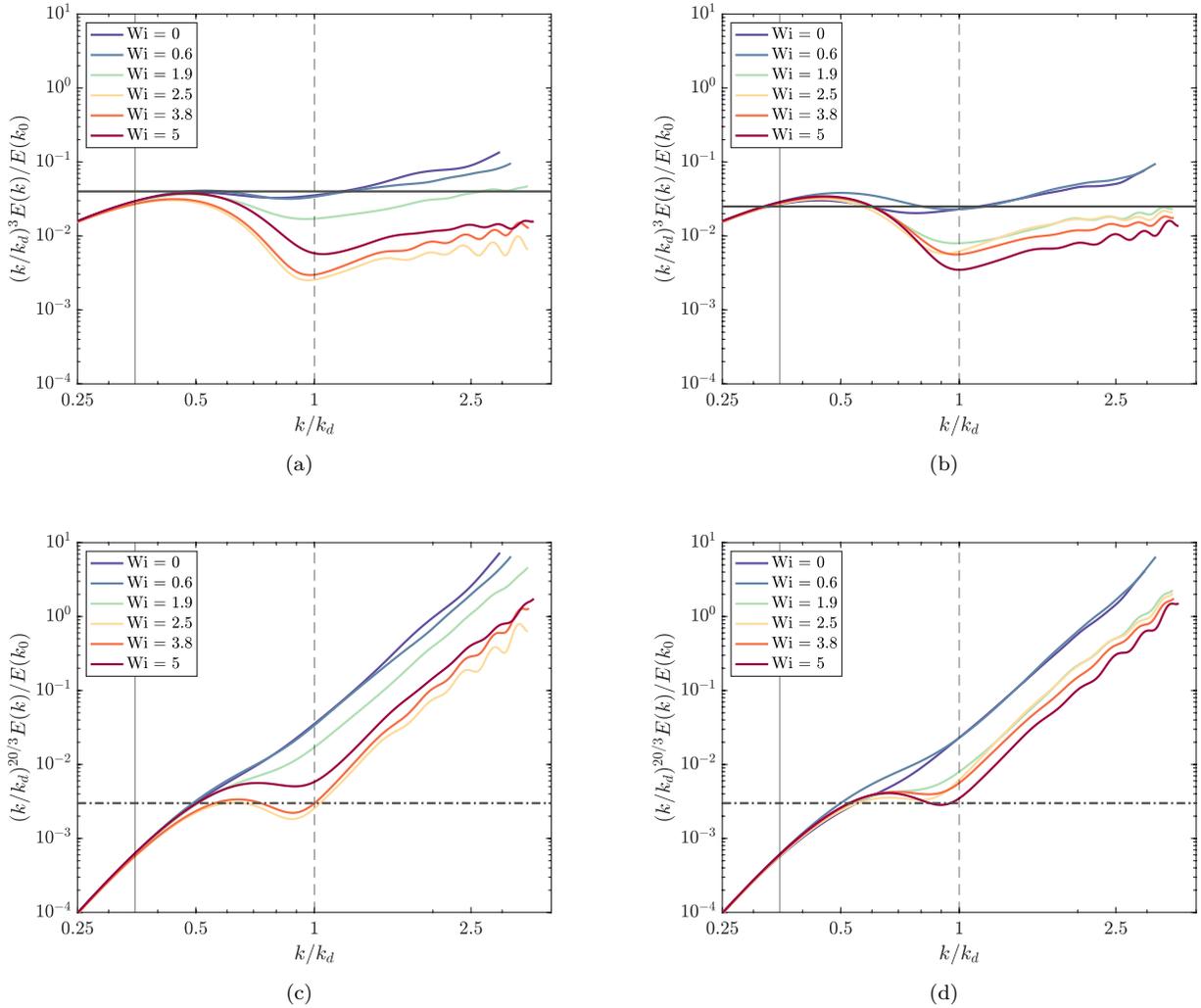

FIG. S9. Compensated (a) horizontal, and (b) vertical spectra of the liquid velocity fluctuations showing the emergence of $k^{-3}$ scaling for a range of Weissenberg numbers at a constant gas volume fraction $\alpha \approx 0.025$. Compensated (c) horizontal, and (d) vertical spectra of the liquid velocity fluctuations showing the emergence of $k^{-20/3}$ scaling for a range of Weissenberg numbers at a constant gas volume fraction $\alpha \approx 0.025$. The abscissa is normalized by the wavenumber corresponding to the bubble diameter, $k_d$. The solid line and the dot-dashed line denote the $k^{-3}$ and $k^{-20/3}$ scaling, respectively.

## EFFECT OF WEISSENBERG NUMBER ON THE DISSIPATION RATE

The turbulent dissipation rate is defined as,

$$\epsilon = 2\nu \langle s_{ij} s_{ij} \rangle, \tag{3}$$

where $\nu$ is the kinematic viscosity and $s_{ij}$ is the fluctuating rate of strain. To estimate the dissipation rate, $\epsilon$, using two velocity components from the PIV data, we use the local isotropic assumption [22],

$$\epsilon = \nu \left\langle 4 \left(\frac{\partial u'_x}{\partial x}\right)^2 + 4 \left(\frac{\partial u'_y}{\partial y}\right)^2 + 3 \left(\frac{\partial u'_x}{\partial y}\right)^2 + 3 \left(\frac{\partial u'_y}{\partial x}\right)^2 + 4 \left(\frac{\partial u'_x}{\partial x} \frac{\partial u'_y}{\partial y}\right) + 6 \left(\frac{\partial u'_x}{\partial y} \frac{\partial u'_y}{\partial x}\right) \right\rangle. \tag{4}$$

Figure S10 illustrates the energy dissipation rate, normalized by its Newtonian counterpart ($\epsilon_0$), as a function of the Weissenberg number (Wi) at a constant gas volume fraction $\alpha \approx 0.025$. For values below the critical Weissenberg number ($\text{Wi}_{\text{cr}} \approx 2.5$), the normalized dissipation rate decreases rapidly. Beyond this threshold, the dissipation rate saturates to 15% of the Newtonian case. This highlights the dissipative characteristics of the polymers, consistent with the previous studies [23–26]. The total average energy injected by the bubbles can be approximated as the work done by the buoyancy force, $P_{\text{total}} \sim \alpha g U_b$. However, only a fraction of this total energy is effectively transferred into the wakes [17, 18]. The inset in Fig. S10 shows the dissipation rate normalized by the total power injected into the system by the bubbles, further emphasizing the role of polymers in the dissipation process.

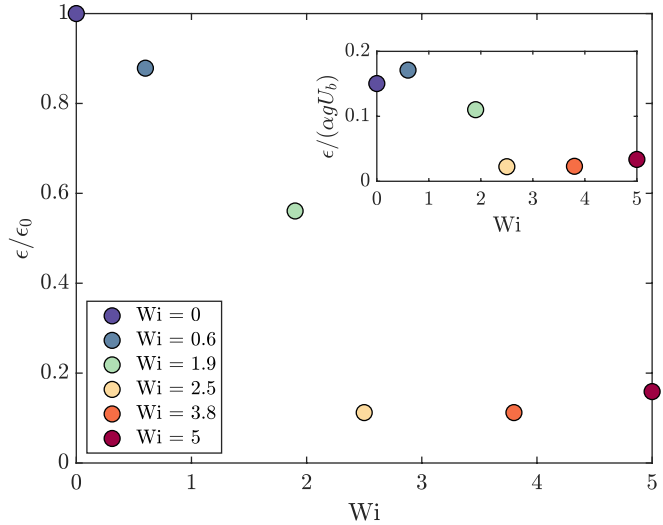

FIG. S10. Dissipation rate, normalized by its Newtonian counterpart, as a function of the Weissenberg number at a constant gas volume fraction of $\alpha \approx 0.025$; Inset shows the dissipation rate normalized by the energy injected through the buoyancy force.

## PROBABILITY DISTRIBUTION OF THE LYAPUNOV EXPONENT

Figure S11 illustrates the skewness in the probability distribution of the Lyapunov exponent, normalized by the corresponding mean, further emphasizing quantitatively that the large-scale flow structures become more coherent.

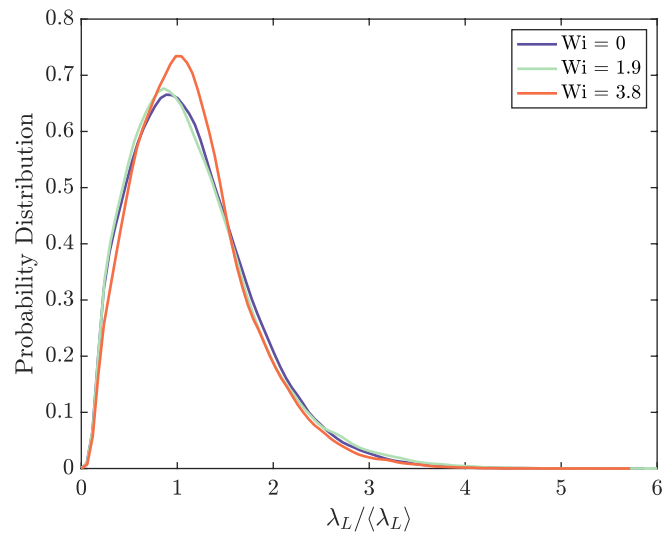

FIG. S11. Probability distribution of the normalized Lyapunov exponent.

[1] J. Martínez-Mercado, C. A. Palacios-Morales, and R. Zenit, Phys. Fluids **19**, 103302 (2007).
[2] S. Mendez-Diaz, J. Serrano-Garcia, R. Zenit, and J. Hernandez-Cordero, Phys. Fluids **25** (2013).
[3] E. Alméras, V. Mathai, D. Lohse, and C. Sun, J. Fluid Mech. **825**, 1091 (2017).
[4] K. C. O. Barbosa, J. D. Cussuol, E. J. Soares, R. M. Andrade, and M. C. Khalil, J. Non-Newtonian Fluid Mech. **310**, 104942 (2022).
[5] W. M. Abed, R. D. Whalley, D. J. C. Dennis, and R. J. Poole, J. Non-Newtonian Fluid Mech. **231**, 68 (2016).
[6] S. Yamani, B. Keshavarz, Y. Raj, T. A. Zaki, G. H. McKinley, and I. Bischofberger, Phys. Rev. Lett. **127**, 074501 (2021).
[7] D. Samanta, Y. Dubief, M. Holzner, C. Schäfer, A. N. Morozov, C. Wagner, and B. Hof, Proc. Natl. Acad. Sci. **110**, 10557 (2013).
[8] J. Dinic, Y. Zhang, L. N. Jimenez, and V. Sharma, ACS Macro Lett. **4**, 804 (2015).
[9] F. Risso and K. Ellingsen, J. Fluid Mech. **453**, 395 (2002).
[10] G. Riboux, F. Risso, and D. Legendre, J. Fluid Mech. **643**, 509 (2010).
[11] V. N. Prakash, J. M. Mercado, L. van Wijngaarden, E. Mancilla, Y. Tagawa, D. Lohse, and C. Sun, J. Fluid Mech. **791**, 174 (2016).
[12] J. H. Lee, H. Kim, J. Lee, and H. Park, Phys. Fluids **33** (2021).
[13] T. Ma, H. Hessenkemper, D. Lucas, and A. D. Bragg, J. Fluid Mech. **936**, A42 (2022).
[14] R. Ran, D. A. Gagnon, A. Morozov, and P. E. Arratia, arXiv preprint arXiv:2111.00068 (2021).
[15] G. Riboux, D. Legendre, and F. Risso, J. Fluid Mech. **719**, 362 (2013).
[16] R. Mei and R. J. Adrian, J. Fluid Mech. **237**, 323 (1992).
[17] F. Risso, Annu. Rev. Fluid Mech. **50**, 25 (2018).
[18] R. Zamansky, F. L. R. De Bonneville, and F. Risso, J. Fluid Mech. **984**, A68 (2024).
[19] A. Cartellier and N. Rivière, Phys. Fluids **13**, 2165 (2001).
[20] A. Varshney and V. Steinberg, Phys. Rev. Fluids **2**, 051301 (2017).
[21] P. C. Valente, C. B. Da Silva, and F. T. Pinho, J. Fluid Mech. **760**, 39 (2014).
[22] D. Xu and J. Chen, Exp. Therm. Fluid Sci. **44**, 662 (2013).
[23] Y. Zhang and H. Xi, Phys. Fluids **34** (2022).
[24] D. Bonn, Y. Couder, P. H. J. Van Dam, and S. Douady, Phys. Rev. E **47**, R28 (1993).
[25] N. T. Ouellette, H. Xu, and E. Bodenschatz, J. Fluid Mech. **629**, 375 (2009).
[26] F. Wang, Y. Zhang, and H. Xi, Exp. Fluids **66**, 1 (2025).